\documentclass[useAMS,usenatbib]{mn2e}

\usepackage{amssymb}
\usepackage{graphicx}
\usepackage{ulem}
\usepackage{color}

\newcommand{\Msun}{\ensuremath{{\rm M}_{\sun}}}
\newcommand{\Lsun}{\ensuremath{{\rm L_{\sun}}}}

\newcommand{\iso}[2]{\hbox{${}^{#1}{\rm #2}$}}

\title[Helium and metal-rich models]{Helium enrichment and Carbon-star Production 
in Metal-rich Populations}
\author[Karakas]{Amanda I. Karakas$^{1,2}$\thanks{E-mail:
amanda.karakas@anu.edu.au}\\
$^{1}$Research School of Astronomy and Astrophysics, 
Australian National University, Canberra, ACT 2611, Australia\\
$^{2}$Kavli Institute for the Physics and Mathematics of the Universe (WPI), 
Todai Institutes for Advanced Study, The University of Tokyo, Japan
}
\begin{document}

\date{}


\maketitle

\label{firstpage}

\begin{abstract}
We present new theoretical stellar evolutionary models of metal-rich 
asymptotic giant branch (AGB) stars.
Stellar models are evolved with initial masses between 1$\Msun$ and 7$\Msun$ 
at $Z=0.007$, and 1$\Msun$ and 8$\Msun$ at $Z=0.014$ (solar) and at $Z=0.03$. 
We evolve models with a canonical helium abundance and with helium 
enriched compositions ($Y=0.30, 0.35, 0.40$) at $Z=0.014$ and $Z=0.03$.
The efficiency of third dredge-up and the mass range of carbon stars 
decreases with an increase in metallicity.  We predict carbon stars form
from initial masses between 1.75--7$\Msun$ at $Z=0.007$ and between 2--4.5$\Msun$
at solar metallicity. At $Z=0.03$ the mass range for C-star production is 
narrowed to 3.25--4$\Msun$. The third dredge-up 
is reduced when the helium content of the model increases owing
to the reduced number of thermal pulses on the AGB. A small increase of $\Delta Y = 0.05$ 
is enough to prevent the formation of C stars at $Z=0.03$, depending on the mass-loss rate, 
whereas at $Z=0.014$,  an increase of $\Delta Y \gtrsim 0.1$ is required to 
prevent the formation of C stars.
We speculate that the probability of finding C stars in a stellar population depends 
as much on the helium abundance as on the metallicity. To explain the paucity 
of C stars in the inner region of M31 we conclude that the observed stars
have $Y \gtrsim 0.35$ or that the stellar metallicity is higher than 
[Fe/H] $\approx 0.1$.
\end{abstract}


\section{Introduction}

The bulge of the Milky Way Galaxy is home to some of the oldest and most metal-rich
stars in our Galaxy. While most of the stars in the bulge are consistent with being
older than 10~Gyr \citep[e.g.,][]{ortolani95,zoccali03,brown10,valenti13}, 
there is evidence for a spread in ages with the 
youngest stars having ages as low as 2~Gyr \citep{bensby13}.  The bulge is also
home to a number of planetary nebulae (PNe) and asymptotic giant branch (AGB) stars
\citep{vanloon03,cole02,gorny04,groen05,uttenthaler07,gorny10,garcia14}.
The AGB stars show evidence of self enrichment through dredge-up processes owing to the 
detection of the heavy element technetium (Tc), although they themselves are not 
carbon rich \citep{uttenthaler07,uttenthaler08}.
Tc is a product of the slow neutron capture process, the $s$-process, which occurs
in the deep interiors of AGB stars and is mixed to the surface by the third dredge-up.
The third dredge-up takes place after a thermal pulse and can occur many times, depending
on the initial mass, metallicity, and H-exhausted core mass \citep[for reviews of AGB evolution and 
nucleosynthesis we refer to][]{busso99,herwig05,karakas14dawes}. However the dominant
product of the third dredge-up is carbon, which is produced as a primary product of
helium burning by the triple-alpha process. The third dredge-up therefore 
mixes carbon and heavy elements to the surface, and is the mechanism for converting
oxygen-rich AGB stars to carbon-rich stars, which have more carbon than oxygen atoms
in their atmospheres, that is, C/O $\ge 1$ \citep[e.g.,][]{wallerstein98}.  

The fact that the bulge AGB stars are rich in Tc means that the third dredge-up 
has occurred, which requires a minimum mass of about $\gtrsim 1.5\Msun$.  
\citet{cole02} found a population of carbon stars
that traced the bar of our Galaxy and speculated that some of these may have 
wandered into the bulge region. The bulge PNe on the other hand show a double 
chemistry, with oxygen bearing molecules found together with polycyclic aromatic 
hydrocarbons, which are carbon-bearing molecules. While the double chemistry phenomena 
may be the result of chemical reactions and not internal nucleosynthesis \citep{guzman11}, 
spectroscopic follow-up observations of bulge PNe find that some of them have 
experienced the third dredge-up \citep{garcia14}. The progenitor
masses of the PNe are not known, with \citet{garcia14} speculating that some of the nebula
evolved from intermediate-mass AGB stars with masses $\ge 4\Msun$.

While the metallicity distribution of stars in the Galactic bulge shows a tail 
down to [Fe/H] $\lesssim -2$\footnote{where we use the standard spectroscopic
notation [Fe/H] = $\log_{10}$(Fe/H)$_{*} - \log_{10}$(Fe/H)$_{\odot}$.}, the mean
metallicity is around solar \citep[e.g.,][]{zoccali08,bensby13}, with 95\%
of stars near the plane having metallicities between $-1 \lesssim$ [Fe/H] 
$\lesssim 0.6$ \citep{bensby13}.
The bulge has been shown to comprise two or three  metallicity components, peaked 
at above solar ([Fe/H] $\approx 0.1-0.3$) and just below solar metallicity with
the metal-rich stars closer to the plane \citep{bab10,hill11,ness13}.
\citet{bensby13} find the younger stellar component resides in the metal-rich bulge. 
The $\alpha$-element content of the metal-rich bulge is approximately solar or
within 0.2~dex of solar for Mg, Si, Ca, and Ti \citep{hill11,bensby13} although
[O/Fe] declines linearly with [Fe/H] \citep[see also][]{cjohnson11,gonzalez11}.
If the AGB stars and PNe are truly in the bulge, they likely evolved from a 
relatively young and metal-rich stellar population with [Fe/H] $\ge 0.0$ and 
[$\alpha$/Fe] $\approx 0$.

Dredge-up in AGB stars is strongly dependent on metallicity as well as mass 
\citep[e.g.,][]{boothroyd88c,karakas02,straniero03}, with lower dredge-up efficiencies
found at solar metallicity compared to the metallicities of the Magellanic Clouds.
However, the studies by \citet*{karakas02} and \citet{straniero03} 
did not include AGB stars with metallicities higher than solar so it is unclear 
how dredge-up efficiencies vary as a function of mass in super-solar metallicity 
AGB stars.  The high initial oxygen abundance of 
metal-rich stars further impedes the formation of carbon stars, because
carbon star formation requires there to be enough carbon atoms to exceed the
now high number of oxygen atoms. These two factors have led to the suggestion
that there is a metallicity ceiling to carbon-star formation, as discussed
by \citet{boyer13} in the context of a paucity of carbon stars in the inner
region of the Andromeda Galaxy (M31). 

Helium enrichment may also be an important factor for stars in the bulges of 
spiral galaxies and in early-type galaxies 
\citep{atlee09,nataf11,chung11,nataf12,rosenfeld12,buell13} 
In order to reconcile the factor $\approx 2$ discrepancy between spectroscopic 
and photometric age determinations of the Galactic bulge main-sequence turnoff, 
\citet{nataf12} suggested that the metal-rich component of the bulge may also be 
helium rich, with helium abundances $\Delta Y \approx 0.1$ up from canonical 
expectations\footnote{where $Y$ is the mass fraction of helium, $X$ the mass 
fraction of hydrogen and $Z$ the mass fraction of metals. $Z$ is also the 
global metallicity of the stellar model.}. \citet{bensby13} note that an increase of helium
by 0.1 does not remove the need for a young and intermediate-age stellar population.
HST photometry has also revealed that some Galactic Globular Clusters also host
helium-rich populations, with helium abundances up to $Y \approx 0.4$ in the case
of $\omega$ Centauri and NGC~2808 \citep[e.g.,][]{norris04,piotto05,dantona05,joo13}.

The effect of helium-enrichment on stellar evolution during the giant branches
is less well understood than its effect on colour-magnitude diagrams. \citet*{karakas14}
studied the effect of helium enrichment on the evolution and nucleosynthesis
of low-mass AGB stars. Helium enrichment was found to severely reduce the
stellar yields expected from low-mass AGB populations at low metallicities by more
than 50\% for some elements. The effect at solar or super-solar metallicities is
not known. Note that the origin of the high helium abundances is unknown, although
it has been speculated that low-metallicity intermediate-mass or super-AGB stars or 
massive stars produced the high helium content of Galactic Globular Clusters
\citep[e.g.,][]{norris04,karakas06b,dercole12}.

The aim of the present study is to provide new detailed stellar evolutionary models
of low and intermediate-mass AGB stars of updated solar metallicity ($Z=0.014$)
and super-solar metallicity ($Z = 0.03$). These models will be used to 
provide the first study of the dependence on mass and helium abundance on the third dredge-up 
at $Z=0.03$, and will be useful for a range of applications including 
synthetic or parametric AGB studies \citep[e.g.,][]{izzard04b,marigo13,buell13}.  
We also map out the mass range of carbon stars at super-solar metallicities from 
detailed stellar evolution models, and examine the effect of helium enrichment 
on the predicted mass range of carbon stars. Given that a large initial helium
abundance truncates the yields of low mass, lower metallicity AGB models \citep{karakas14}, 
it is reasonable to expect that enhanced helium may inhibit carbon star 
production at higher metallicities.
In this paper we first introduce the stellar evolutionary models in \S\ref{sec:models}
including a discussion of the initial helium abundance. We then present
the results of the new stellar models in \S\ref{sec:results} including a comparison
to other studies, discuss the major uncertainties affecting the results in \S\ref{sec:uncert},
and finish with a discussion and conclusions in \S\ref{sec:conclude}.

\section{Stellar Evolutionary Models} \label{sec:models}

In this study we evolve stellar models of mass 1$\Msun$ to 7$\Msun$ 
with a global metallicity of $Z=0.007$, and 1$\Msun$ to 8$\Msun$ with 
metallicities of $Z= 0.014$ (solar) and $Z=0.03$. The full grid of masses at each metallicity is given
in Table~\ref{tab:models}. Models are evolved from the pre-main sequence to the 
tip of the AGB. The maximum masses at each metallicity experience off-centre carbon ignition but the
carbon burning does not reach the centre. These are CO(Ne) core AGB stars according
to the definitions given in \citet{karakas14dawes} and are not true super-AGB stars,
which have O-Ne cores \citep[e.g.,][]{siess10,doherty10}.
The metallicities were chosen so we include models of solar metallicity and a metallicity
that is approximately a factor of two more metal-poor and metal-rich than solar.

The metallicities of the models are representative of disc metallicities, according to 
iron abundances derived from stars
\citep{edvardsson93,casagrande11,bensby14,recio-blanco14}, and abundances of oxygen and 
zinc in planetary nebulae, which are tracers of Galactic disc metallicities 
\citep[e.g.,][]{stasinska98,stanghellini10,csmith14}.
The solar and metal-rich models are appropriate for comparison to stars and planetary nebulae in 
the metal-rich bulge of our Milky Way Galaxy \citep{bensby13,ness13}, and also for comparison to 
AGB stars found in the inner regions of spiral galaxies such as M31 \citep{saglia10,boyer13}. 
The lower metallicity models of $Z=0.007$ (or [Fe/H] $\approx -0.3$) are similar to the 
the metallicity of thick disc \citep[e.g.,][]{reddy06,recio-blanco14} or the peak
metallicity of the Large Magellanic Cloud \citep[LMC,][]{cole05}, 
and will mostly be used here for comparison to the metal-rich models.

The input physics used in the stellar evolutionary sequences is exactly the same
as described in \citet{karakas14}. The initial composition of C, N, and O are scaled solar
in the $Z=0.03$ and $Z=0.007$ models and solar in the $Z = 0.014$ models, where the solar 
abundances are from \citet{asplund09}. We choose to use the \citet{asplund09} 
abundances for comparison to other recent solar-metallicity stellar evolution models 
\citep[e.g., by][who adopt $Z=0.014$ or a value close to that]{cristallo11,ekstrom12}.
Furthermore, we show in \S\ref{sec:uncert} that the stellar evolution
calculations are not dependent on the choice of solar abundances and that
adopting the say, \citet{lodders09} solar abundances, which have $Z_{\odot} = 0.0153$, does
not change the results for a 3$\Msun$ model. In the $Z=0.007$ models we assume a scaled-solar
composition, noting that there is little or no $\alpha$-enhancement in the Galactic thin disc
at [Fe/H] $=-0.3$ and only a mild $\alpha$-enhancement in the thick disc at these metallicities
\citep[e.g., Figs.~12 and 13 from][]{reddy06}.

The initial helium abundance is varied 
in the solar and metal-rich models ($Z=0.014, 0.03$, respectively) and is described in more
detail below.    We assume no mass loss on the red giant branch (RGB) and use the 
\citet{vw93} mass-loss formulation on the AGB. While the assumption of no mass loss on 
the RGB is  an incorrect assumption, the Kepler results by \citet{miglio12} 
suggest that the mass-loss rates in metal-rich open cluster giant stars are less 
than predicted by Reimer's type mass-loss prescriptions with $\eta \approx 0.4$.
We use the C and N-rich low-temperature opacity tables from \citet{marigo09}, which
are based on the solar composition of \citet{lodders03}. The OPAL tables use the
same initial composition as the low-temperature tables for consistency. We note that
the initial solar $Z$ is slightly lower in the opacity tables than we assume here in
the stellar evolutionary calculations ($Z = 0.01321$ in the opacity tables compared to 0.014) 
but the difference is small ($\Delta Z = 0.00079$).

Convection is approximated using the Mixing-length Theory with a mixing-length 
parameter of $\alpha = 1.86$ in all calculations. No convective overshoot is applied 
although we use the algorithm described by \citet{lattanzio86} to search for a neutrally 
stable point for the border between convective and radiative zones. This has been shown to increase 
the amount of third dredge-up relative to models that set the position of the convective border
according to the formal Schwarzschild boundary \citep{frost96,mowlavi99a}.  \citet*{kamath12}
found that this scheme was not able to reproduce the observations of AGB stars in Magellanic
Cloud clusters and further mixing was required. \citet{kamath12} required a large amount of 
convective overshoot (up to 3 pressure scale heights)  at the base of the convective 
envelope during third dredge-up in order to match the O-rich to C-rich transition luminosity 
of the cluster AGB stars.  Here we ignore further mixing until \S\ref{sec:compare}, noting
this is a considerable uncertainty to the lower mass limit for carbon 
star production. 

\subsection{The initial helium abundance}

The primordial helium abundance, $Y_{\rm p}$, is a firm lower limit to the initial abundance 
of helium of the first stars in the Universe. For other generations of stars, the helium
abundance has been steadily increasing as a result of stellar nucleosynthesis and is 
determined according to:
\begin{equation}
 Y =  \frac{\Delta Y}{\Delta Z} Z + Y_{\rm p},
\end{equation}
where $\Delta Y/\Delta Z$ is the rate of helium production and is typically 
expressed relative to the change in metallicity, $Z$. Both
$Y_{\rm p}$ and the gradient can be estimated from observations. \citet{aver13} estimate 
a value for $Y_{\rm p} = 0.2485 \pm 0.0002$, based on the Planck determination of the baryon 
density and using the most recent He I emissivities based on improved photoionization 
cross-sections from \citet{porter12}, and a re-analysis of the observations by \citet{izotov07}.
The slope, $\Delta Y/\Delta Z$, has been estimated to be between 1 to 10 
\citep{chiappini02,balser06,gennaro10,portinari10}, with \citet{balser06} finding a value of 
$1.41 \pm 0.62$ in the Galaxy, consistent with standard chemical evolution models 
\citep[e.g.,][]{chiappini02}.  \citet{izotov07} estimate a value closer to 3 (2.94 or 2.88, depending 
upon their choice of He I emissivities) for low metallicity extra-galactic HII regions, 
whereas \citet{casagrande07} estimate $\Delta Y/\Delta Z$ to be $2.1 \pm 0.9$ around and 
above solar metallicity. 

In the models with global metallicities set to $Z=0.014$ and $Z=0.03$, we investigate the 
effect of varying the initial helium abundance on the stellar evolutionary sequences, and 
in particular, on the evolution during the AGB. In the $Z=0.014$ models we set our canonical 
value $Y = 0.28$ and in the $Z=0.03$ models we set the canonical $Y=0.30$ 
(where $X+Y+Z=1$, noting that when we vary $Y$ we keep $Z$ constant, which means that the
hydrogen abundance, $X$, also varies). We set $Y=0.26$ in all the $Z=0.007$ models.
We then evolve a series of models with masses between 2$\Msun$ and 5$\Msun$ at $Z=0.014$ and
$Z= 0.03$ with helium abundances shown in Table~\ref{tab:he}.  The mass range of models 
was chosen to investigate the effect of helium enrichment on the production of carbon stars.

If we take $Y_{\rm p} = 0.2485$ \citep{aver13} and $\Delta Y/\Delta Z = 2.1$ \citep{casagrande07},
then at $Z=0.007, 0.014$ and $Z = 0.03$ we get $Y = 0.2632, 0.2779$, and $Y = 0.3115$, respectively.
These are close to our chosen canonical values at our metallicities although it suggests that 
our choice of $Y = 0.28$ is a bit high for solar metallicity, which is motivation to run a few models
with a lower value of $Y=0.26$. For the $Z=0.03$ models, the standard helium abundance of $Y=0.30$
is a bit low for metal-rich stars, which is motivation for calculating a few stellar models with 
initial helium  of $Y=0.32$. Helium-enriched models at both metallicities include those calculated
with $Y =0.35$ and $Y = 0.40$. Note that at $Z=0.014$, an initial helium of $Y=0.35$ or 0.40 implies
a slope of $\Delta Y/\Delta Z = 7.25$ and 10.82, respectively, whereas at $Z=0.03$ the slope is
3.38 and 5.05 for $Y = 0.35$ and 0.40. 

\section{Results} \label{sec:results}
 
\begin{table*}
\begin{center}
\caption{Stellar models calculated with a canonical helium composition. The luminosity
is in the format $n(m)$ where $= n \times 10^{m}\Lsun$.}
\label{tab:models}
\begin{tabular}{lccccccccccccccc} \hline \hline
Mass & SDU & HBB & TDU & \#TP & C/O$_{\rm f}$ & $\lambda_{\rm max}$ & $M_{\rm c}(1)$ 
& $M_{\rm c}^{\rm min}$ & $T_{\rm bce}^{\rm max}$ & $L_{\rm agb}^{\rm max}$ & 
$\tau_{\rm stellar}$ & $\tau_{\rm agb}$ & $\tau_{\rm tpagb}$ & $\tau_{\rm c}$  & 
$\tau_{\rm c}/\tau_{\rm agb}$ \\
($\Msun$) &  &     &  &    &     &    & ($\Msun$)     & ($\Msun$) & (MK) & ($\Lsun$) & 
 (Myr) & (Myr) & (Myr)  &  (Myr) & \\ \hline
\multicolumn{14}{c}{$Z=0.007$, $Y=0.26$ models.} \\ \hline
1.00  & No & No & No  & 15 & 0.463 & 0.00 & 0.537 & -- & 1.75 & 5.69(3) &
10065 & 20.66 & 1.875 & -- & -- \\
1.25  & No & No & No  & 16 & 0.403 & 0.00 & 0.545 & -- & 2.28 & 6.99(3) &
4526 & 19.23 & 1.921 & -- & -- \\
1.50  & No & No & Yes & 19 & 0.509 & 0.10 & 0.549 & 0.636 & 2.78 & 8.55(3) &
2451 & 18.70 & 2.112 & -- & -- \\
1.75  & No & No & Yes & 20 & 1.566 & 0.49 & 0.554 & 0.638 & 3.19 & 9.98(3) &
1535 & 17.28 & 2.155 & 0.137 & 0.008 \\
1.90  & No & No & Yes & 21 & 2.475 & 0.52 & 0.552 & 0.615 & 3.45 & 9.64(3) &
1216 & 17.69 & 2.193 & 0.317 & 0.018 \\
2.10  & No & No & Yes & 24 & 3.397 & 0.60 & 0.538 & 0.617 & 3.63 & 9.44(3) &
968.5 & 21.13 & 2.801 & 0.494 & 0.023 \\
2.25  & No & No & Yes & 27 & 4.254 & 0.68 & 0.532 & 0.608 & 3.92 & 1.06(4) &
900.2 & 23.72 & 3.185 & 0.662 & 0.028 \\
2.50  & No & No & Yes & 27 & 5.389 & 0.76 & 0.555 & 0.609 & 4.64 & 1.14(4) &
687.3 & 17.06 & 2.621 & 0.897 & 0.053 \\
3.00  & No & No & Yes & 22 & 6.820 & 0.87 & 0.649 & 0.658 & 7.25 & 1.36(4) &
394.8 & 8.239 & 1.361 & 0.935 & 0.114 \\
3.50  & No & No & Yes & 21 & 5.812 & 0.97 & 0.746 & 0.750 & 17.8 & 1.72(4) &
256.8 & 4.770 & 0.670 & 0.503 & 0.105 \\
4.00  & Yes & No & Yes & 24 & 4.144 & 0.98 & 0.837 & 0.839 & 46.1 & 2.28(4) &
181.7 & 2.984 & 0.350 & 0.251 & 0.084 \\
4.25  & Yes & Yes & Yes & 28 & 3.563 & 0.97 & 0.846 & 0.849 & 56.5 & 2.47(4) &
155.1 & 2.625 & 0.358 & 0.248 & 0.095 \\
4.50  & Yes & Yes & Yes & 50 & 1.118 & 0.96 & 0.856 & 0.858 & 76.4 & 3.19(4) &
134.9 & 2.352 & 0.543 & 0.014  & 0.006 \\
5.00  & Yes & Yes & Yes & 59 & 1.881 & 0.95 & 0.876 & 0.878 & 82.9 & 3.69(4) &
104.3 & 1.736 & 0.500 & 0.023 & 0.013 \\
5.50  & Yes & Yes & Yes & 67 & 1.947 & 0.93 & 0.902 & 0.903 & 86.8 & 4.17(4) &
83.54 & 1.270 & 0.408 & 0.018 & 0.014 \\
6.00  & Yes & Yes & Yes & 64 & 2.075 & 0.92 & 0.940 & 0.941 & 92.0 & 4.75(4) &
68.89 & 0.846 & 0.259 & 0.017 & 0.021 \\
7.00  & Yes & Yes & Yes & 61 & 2.040 & 0.90 & 1.030 & 1.031 & 102 & 6.15(4) &
48.93 & 0.117 & 0.096 & 0.004 & 0.038 \\ \hline
\multicolumn{14}{c}{$Z=0.014$, $Y=0.28$ models.} \\ \hline
1.00  & No  & No & No  & 12 & 0.469 & 0.00 & 0.546 & -- & 1.77 & 4.81(3) &
12186 & 21.34 & 1.100 & -- & -- \\
1.25  & No  & No & No  & 14 & 0.412 & 0.00 & 0.550 & -- & 2.39 & 6.00(3) &
5372  & 19.55 & 1.286 &  -- & -- \\
1.50  & No  & No & No  & 17 & 0.380 & 0.00 & 0.555 & -- & 2.76 & 7.43(3) &
2882  & 18.26 & 1.508 & -- & -- \\
1.75  & No  & No & No  & 20 & 0.354 & 0.00 & 0.557 & -- & 3.10 & 8.74(3) &
1756  & 18.55 & 1.702 & -- &  -- \\
2.00  & No  & No & Yes & 25 & 1.233 & 0.50 & 0.531 & 0.616 & 2.80 & 1.03(4) &
1186  & 18.22 & 2.612 & 0.138 & 0.008 \\
2.25  & No  & No & Yes & 32 & 1.331 & 0.62 & 0.536 & 0.631 & 4.76 & 1.12(4) &
1015  & 26.79 & 2.746 & 0.114 & 0.004 \\
2.50  & No  & No & Yes & 31 & 1.744 & 0.72 & 0.546 & 0.638 & 5.22 & 1.18(4) &
770.2 & 22.40 & 2.534 & 0.294 & 0.013 \\
2.75  & No  & No & Yes & 30 & 2.421 & 0.77 & 0.567 & 0.634 & 5.48 & 1.24(4) &
587.5 & 16.89 & 2.141 & 0.459 & 0.027 \\
3.00  & No  & No & Yes & 28 & 2.700 & 0.80 & 0.598 & 0.641 & 6.30 & 1.30(4) &
453.6 & 12.19 & 1.706 & 0.604 & 0.050 \\
3.25  & No  & No & Yes & 25 & 2.999 & 0.81 & 0.644 & 0.664 & 7.77 & 1.41(4) &
355.5 & 8.703 & 1.243 & 0.627 & 0.072 \\
3.50  & No  & No & Yes & 24 & 2.576 & 0.90 & 0.691 & 0.700 & 11.9 & 1.57(4) &
 282.0 & 6.384 & 0.869 & 0.473 & 0.074 \\
4.00  & No  & No & Yes & 23 & 1.939 & 0.96 & 0.797 & 0.800 & 34.8 & 2.08(4) &
192.7 & 3.899 & 0.392 & 0.168 & 0.043 \\
4.50  & Yes & Yes & Yes & 31 & 1.287 & 0.96 & 0.847 & 0.850 & 63.5 & 2.73(4) &
141.0 & 2.607 & 0.338 & 0.028 & 0.011 \\
5.00  & Yes & Yes & Yes & 41 & 0.844 & 0.95 & 0.863 & 0.866 & 75.4 & 3.06(4) &
108.2 & 1.996 & 0.352 & --   & --   \\
5.50  & Yes & Yes & Yes & 49 & 0.854 & 0.94 & 0.883 & 0.884 & 81.4 & 3.49(4) &
85.07 & 1.558 & 0.334 & --   & --  \\
6.00  & Yes & Yes & Yes & 51 & 0.894 & 0.93 & 0.905 & 0.907 & 85.5 & 3.96(4) &
69.85 & 1.160 & 0.282 & --  & --   \\
7.00  & Yes & Yes & Yes & 56 & 0.769 & 0.92 & 0.962 & 0.964 & 92.4 & 4.99(4) &
48.44 & 0.709 & 0.166 & --   & --   \\
8.00  & Yes & Yes & Yes & 67 & 0.570 & 0.87 & 1.052 & 1.053 & 100 & 6.37(4) & 
36.52 & 0.435 & 0.086 & --   & --   \\ \hline
\multicolumn{14}{c}{$Z=0.03$, $Y=0.30$ models.} \\ \hline
1.00  & No  & No & No  & 6 & 0.478 & 0.00 & 0.580 & -- & 1.76 & 3.99(3) & 
16164 & 22.84 & 0.420 &  -- & -- \\
1.25  & No  & No & No  & 10 & 0.423 & 0.00 & 0.560 & -- & 2.42 & 5.10(3) &
7004  & 21.86 & 0.679 &  -- & -- \\
1.50  & No  & No & No  & 14 & 0.388 & 0.00 & 0.561 & -- & 2.75 & 6.26(3) &
3655  & 20.71 & 0.922 &  -- & -- \\
1.75  & No  & No & No  & 17 & 0.362 & 0.00 & 0.563 & -- & 3.05 & 7.33(3) &
2183  & 18.78 & 1.119 &  -- & -- \\
2.00  & No  & No & No  & 22 & 0.355 & 0.00 & 0.559 & -- & 3.37 & 8.45(3) &
1450  & 20.65 & 1.403 & -- & -- \\
2.25  & No  & No & No  & 28 & 0.349 & 0.00 & 0.547 & -- & 3.76 & 9.66(3) &
1161  & 28.12 & 1.895 &  -- & -- \\
2.50  & No  & No & Yes & 31 & 0.359 & 0.09 & 0.551 & 0.668 & 4.27 & 1.10(4) &
914.2 & 25.85 & 1.984 & -- & -- \\
2.75  & No  & No & Yes & 33 & 0.523 & 0.44 & 0.564 & 0.666 & 5.09 & 1.22(4) &
695.0 & 20.83 & 1.768 & --   & -- \\
3.00  & No  & No & Yes & 33 & 0.911 & 0.70 & 0.580 & 0.665 & 6.20 & 1.33(4) &
532.3 & 16.79 & 1.621 & --   & --  \\
3.25  & No  & No & Yes & 32 & 1.307 & 0.78 & 0.611 & 0.667 & 7.91 & 1.46(4) &
417.1 & 11.83 & 1.346 & 0.141 & 0.012 \\
3.50  & No  & No & Yes & 33 & 1.523 & 0.84 & 0.646 & 0.677 & 12.7 & 1.59(4) &
327.8 & 9.610 & 1.161 & 0.254 & 0.026 \\
3.75  & No  & No & Yes & 32 & 1.497 & 0.86 & 0.687 & 0.705 & 16.0 & 1.71(4) &
264.1 & 7.522 & 0.889 & 0.246 & 0.033 \\
4.00  & No  & No & Yes & 24 & 1.226 & 0.91 & 0.741 & 0.751 & 19.0 & 1.86(4) &
219.2 & 5.197 & 0.496 & 0.060 & 0.012 \\
4.25  & No  & No & Yes & 19 & 0.894 & 0.94 & 0.790 & 0.796 & 24.4 & 2.05(4) &
182.1 & 4.108 & 0.260 & --   & --   \\
4.50  & Yes & No & Yes & 20 & 0.786 & 0.93 & 0.832 & 0.836 & 31.9 & 2.26(4) &
154.8 & 3.299 & 0.189 & --  & --   \\
4.75  & Yes & No & Yes & 21 & 0.838 & 0.94 & 0.843 & 0.846 & 42.6 & 2.41(4)  &
132.2 & 2.668 & 0.178 & --   & --   \\
5.00  & Yes & Yes & Yes & 26 & 0.885 & 0.94 & 0.851 & 0.855 & 54.2 & 2.59(4) &
115.7 & 2.306 & 0.202 & --   & --   \\
5.50  & Yes & Yes & Yes & 31 & 0.733 & 0.94 & 0.869 & 0.873 & 64.7 & 2.97(4) &
90.08 & 1.662 & 0.191 & -- & -- \\
6.00  & Yes & Yes & Yes & 33 & 0.604 & 0.93 & 0.891 & 0.894 & 71.2 & 3.35(4) &
72.55 & 1.218 & 0.176 & --  & --   \\
7.00  & Yes & Yes & Yes & 43 & 0.349 & 0.90 & 0.948 & 0.951 & 82.7 & 4.22(4) &
50.14 & 0.708 & 0.120 & --  & --   \\
8.00  & Yes & Yes & Yes & 63 & 0.290 & 0.85 & 1.041 & 1.043 & 94.0 & 5.73(4) &
36.64 & 0.434 & 0.078 & --  & --   \\
\hline \hline
\end{tabular}
\medskip\\
\end{center}
\end{table*}

\begin{table*}
\begin{center}
\caption{Stellar models calculated with variable helium compositions.}
\label{tab:he}
\begin{tabular}{lccccccccccccccc} \hline \hline
Mass & SDU & HBB & TDU & \#TP & C/O$_{\rm f}$ & $\lambda_{\rm max}$ & $M_{\rm c}(1)$ 
& $M_{\rm c}^{\rm min}$ & $T_{\rm bce}^{\rm max}$ & $L_{\rm agb}^{\rm max}$  & 
$\tau_{\rm stellar}$ & $\tau_{\rm agb}$ & $\tau_{\rm tpagb}$ & $\tau_{\rm c}$  & 
$\tau_{\rm c}/\tau_{\rm agb}$ \\
($\Msun$) &     &     &       &    &   &  & ($\Msun$)     & ($\Msun$) & (MK) & ($\Lsun$)  & 
 (Myr) & (Myr) & (Myr)  &  (Myr) & \\ \hline 
\hline
\multicolumn{14}{c}{$Z=0.014$, $Y=0.26$ models.} \\ 
2.0  & No & No  & Yes & 24 & 1.179 & 0.46 & 0.544 & 0.631 & 3.65 & 1.01(4) &
1324 & 20.88 & 2.408 & 0.061 & 0.003 \\
2.25 & No & No  & Yes & 30 & 1.436 & 0.64 & 0.529 & 0.624 & 4.42 & 1.05(4) &
1065 & 27.07 & 3.045 & 0.212 & 0.008 \\
2.5  & No & No  & Yes & 30 & 1.850 & 0.73 & 0.535 & 0.625 & 4.61 & 1.11(4) &
858.8 & 25.83 & 2.990 & 0.355 & 0.014 \\
3.0  & No & No  & Yes & 29 & 3.037 & 0.81 & 0.579 & 0.632 & 5.87 & 1.26(4) &
509.7 & 11.47 & 2.163 & 0.774 & 0.054 \\
4.0  & No & No  & Yes & 24 & 2.525 & 0.96 & 0.770 & 0.774 & 24.2 & 1.91(4) &
213.8 & 4.426 & 0.536 & 0.284 & 0.064 \\
4.5  & Yes & Yes & Yes & 32 & 1.218 & 0.96 & 0.842 & 0.844 & 63.2 & 2.67(4) &
153.8 & 2.968 & 0.408 & 0.029 & 0.010 \\
\multicolumn{14}{c}{$Z=0.014$, $Y=0.30$ models.} \\ 
2.0  & No & No  & Yes & 27 & 0.416 & 0.16 & 0.555 & 0.654 & 3.59 & 1.03(4) &
1101 & 19.06 & 1.882 &  -- &  -- \\
2.25 & No & No  & Yes & 33 & 1.289 & 0.64 & 0.548 & 0.655 & 4.78 & 1.20(4) &
912.9 & 23.72 & 2.322 & 0.048 & 0.002 \\
2.5  & No & No  & Yes & 32 & 1.616 & 0.69 & 0.561 & 0.653 & 5.62 & 1.26(4) &
691.4 & 18.47 & 2.093 & 0.188 & 0.010 \\
3.0  & No & No  & Yes & 26 & 2.313 & 0.79 & 0.619 & 0.657 & 6.88 & 1.38(4) &
400.6 & 10.29 & 1.289 & 0.410 & 0.040 \\
4.0  & Yes & No  & Yes & 23 & 1.750 & 0.95 & 0.826 & 0.828 & 47.6 & 2.32(4) &
173.3 & 3.277 & 0.291 & 0.108 & 0.033 \\
4.5  & Yes & Yes & Yes & 29 & 1.206 & 0.94 & 0.854 & 0.856 & 64.1 & 2.80(4) &
127.8 & 2.364 & 0.268 & 0.024 & 0.010 \\
\multicolumn{14}{c}{$Z=0.014$, $Y=0.35$ models.} \\ 
2.0  & No & No  & No  & 27 & 0.340 & 0.00 & 0.579 &  --   & 3.60 & 1.11(4) &
906.9 & 16.65 & 1.283 & --  & -- \\
2.25 & No & No  & Yes & 30 & 0.432 & 0.21 & 0.590 & 0.686 & 4.40 & 1.26(4) &
694.3 & 14.36 & 1.239 & --   & -- \\
2.5  & No & No  & Yes & 31 & 1.209 & 0.60 & 0.612 & 0.680 & 5.99 & 1.43(4) &
517.7 & 11.47 & 1.095 & 0.055 & 0.005 \\
3.0  & No & No  & Yes & 24 & 1.708 & 0.79 & 0.686 & 0.711 & 12.2 & 1.64(4) &
299.8 & 6.729 & 0.589 & 0.153 & 0.023 \\
3.5  & No & No  & Yes & 22 & 1.687 & 0.94 & 0.796 & 0.802 & 30.0 & 2.16(4) &
192.6 & 3.740 & 0.266 & 0.086 & 0.023 \\
4.0  & Yes & Yes & Yes & 23 & 1.275 & 0.93 & 0.855 & 0.859 & 50.4 & 2.57(4) &
134.1 & 2.468 & 0.159 & 0.028 & 0.011 \\
4.5  & Yes & Yes & Yes & 30 & 1.087 & 0.92 & 0.874 & 0.877 & 66.1 & 3.02(4) &
100.3 & 1.690 & 0.162 & 0.007 & 0.004 \\
5.0  & Yes & Yes & Yes & 38 & 0.800 & 0.91 & 0.895 & 0.899 & 78.2 & 3.46(4) &
77.37 & 1.249 & 0.157 & --   & -- \\
\multicolumn{14}{c}{$Z=0.014$, $Y=0.40$ models.} \\ 
2.0  & No & No  & No  & 20 & 0.323 & 0.00 & 0.632 & --    & 3.60 & 1.19(4) &
706.8 & 10.23 & 0.540 & --  & -- \\
2.25 & No & No  & No  & 24 & 0.333 & 0.00 & 0.644 &  --   & 4.41 & 1.36(4) &
511.1 & 9.674 & 0.575 & --   & -- \\ 
2.5  & No & No  & Yes & 22 & 0.357 & 0.13 & 0.677 & 0.729 & 5.66 & 1.52(4) &
378.4 & 7.313 & 0.427 & --   & -- \\
3.0  & No & No  & Yes & 17 & 0.755 & 0.56 & 0.773 & 0.787 & 12.7 & 1.93(4) &
224.7 & 4.058 & 0.169 & --   & -- \\
4.0  & Yes & Yes & Yes & 26 & 0.994 & 0.80 & 0.879 & 0.884 & 53.3 & 2.80(4) &
104.5 & 1.660 & 0.101 & --   & -- \\
4.5  & Yes & Yes & Yes & 34 & 0.727 & 0.79 & 0.905 & 0.912 & 69.6 & 3.36(4) &
77.85 & 1.111 & 0.092 & -- & -- \\ \hline
\multicolumn{14}{c}{$Z=0.03$, $Y=0.32$ models.} \\ 
3.0  & No & No & Yes & 31 & 0.667 & 0.60 & 0.600 & 0.674 & 6.20 & 1.37(4) &
 532.3 & 16.76 & 1.585 & -- & -- \\
3.5  & No & No & Yes & 28 & 1.115 & 0.81 & 0.669 & 0.698 & 11.5 & 1.61(4) &
290.8 & 8.065 & 0.761 & 0.050 & 0.006 \\
4.0  & No & No & Yes & 19 & 0.838 & 0.87 & 0.771 & 0.779 & 19.0 & 1.94(4) &
194.2 & 4.389 & 0.269 & --  & -- \\
5.0  & Yes & Yes & Yes & 25 & 0.799 & 0.92 & 0.859 & 0.863 & 55.1 & 2.67(4) &
104.1 & 2.056 & 0.160 & --   & -- \\
\multicolumn{14}{c}{$Z=0.03$, $Y=0.35$ models.} \\ 
3.0  & No & No & Yes & 28 & 0.461 & 0.39 & 0.632 & 0.697 & 6.10 & 1.44(4) &
387.0 & 10.95 & 0.783 & -- & -- \\
3.25 & No & No & Yes & 23 & 0.615 & 0.58 & 0.680 & 0.713 & 7.22 & 1.48(4) &
307.0 & 6.986 & 0.465 & -- & -- \\
3.5  & No & No & Yes & 21 & 0.699 & 0.72 & 0.717 & 0.736 & 11.4 & 1.72(4) &
242.3 & 6.023 & 0.341 & --   & -- \\
4.0  & Yes & No & Yes & 14 & 0.521 & 0.75 & 0.822 & 0.828 & 18.9 & 2.15(4) &
163.1 & 3.463 & 0.106 & --  & -- \\
4.25 & Yes & No & Yes & 18 & 0.640 & 0.85 & 0.842 & 0.848 & 24.8 & 2.34(4) &
138.1 & 2.876 & 0.118 & --  & -- \\
4.5  & Yes & No & Yes & 20 & 0.632 & 0.85 & 0.852 & 0.857 & 33.3 & 2.45(4) &
118.6 & 2.317 & 0.118 & -- & -- \\ 
4.75 & Yes & No & Yes & 24 & 0.673 & 0.86 & 0.861 & 0.867 & 44.8 & 2.61(4) &
101.9 & 1.971 & 0.126 & --  & -- \\
5.0  & Yes & Yes & Yes & 27 & 0.697 & 0.88 & 0.870 & 0.875 & 56.3 & 2.80(4) &
89.68 & 1.666 & 0.130 & --   & -- \\
\multicolumn{14}{c}{$Z=0.03$, $Y=0.40$ models.} \\ 
3.0  & No & No & No  & 16 & 0.349 & 0.00 & 0.713 & --    & 6.00 & 1.60(4) &
283.0 & 6.141 & 0.223 & -- & -- \\
3.5  & No & No & No & 13 & 0.344 &  0.00 & 0.813 & -- & 10.3 & 2.09(4) &
180.9 & 3.540 & 0.080 & -- & -- \\
4.0  & Yes & No & Yes & 14 & 0.384 & 0.31 & 0.855 & 0.863 & 19.7 & 2.41(4) &
123.6 & 2.356 & 0.060 & --  & -- \\
5.0  & Yes & Yes & Yes & 33 & 0.537 & 0.66 & 0.894 & 0.904 & 59.2 & 3.09(4) &
68.72 & 1.242 & 0.091 & --  & --  \\
\hline \hline
\end{tabular}
\medskip\\
\end{center}
\end{table*}

In Tables~\ref{tab:models} and~\ref{tab:he} we present the list of stellar models calculated for this study. 
We note if the second (SDU) or third dredge-up (TDU) occur and if hot bottom burning (HBB) is active. 
We include the total number of thermal pulses (\#TP), the final C/O ratio in the envelope 
(by number), the maximum third dredge-up efficiency, $\lambda_{\rm max}$, the H-exhausted core mass 
(hereafter core mass) at the first thermal pulse, $M_{\rm c}(1)$, the core mass at the 
first third dredge-up episode, $M_{\rm c}^{\rm min}$, the maximum temperature at the base of 
the convective envelope, $T_{\rm bce}^{\rm max}$, the maximum luminosity on the TP-AGB, 
$L_{\rm agb}^{\rm max}$, the total stellar lifetime, $\tau_{\rm stellar}$, the lifetime on the
AGB including early-AGB, $\tau_{\rm agb}$, the lifetime on the TP-AGB, $\tau_{\rm tpagb}$, the
lifetime during the C-rich phase, $\tau_{\rm c}$, and the ratio between the C-rich lifetime
and the lifetime on the AGB, ($\tau_{\rm c}$)/($\tau_{\rm agb}$).

The third dredge-up efficiency is defined according  to 
$\lambda  = \Delta M_{\rm dredge}/\Delta M_{\rm core}$, 
where $\lambda$ is the third dredge-up efficiency parameter, $\Delta M_{\rm dredge}$ is the mass mixed 
into the envelope, and $\Delta M_{\rm core}$ is the amount by which the H-exhausted core increases over 
the previous interpulse phase. Masses and luminosities are in solar units, temperatures in
$10^{6}$~K, and ages in Myr.

We define low-mass stars as those that experience the core helium flash and intermediate-mass
stars as those that ignite helium under non-degenerate conditions. The maximum mass for the
core helium flash is 2$\Msun$ at $Z=0.007$, 2.25$\Msun$ at $Z=0.014$ and 2.5$\Msun$ at $Z=0.03$. 
Note that we do not include a model of 2$\Msun$ at $Z = 0.007$ owing to convergence difficulties
during core He ignition. All models experience the FDU. 
The SDU requires a minimum H-exhausted core mass of 0.8$\Msun$ on the early AGB and this is 
satisfied by models of 4.5$\Msun$ at $Z=0.014$ and 5$\Msun$ at $Z = 0.03$.
At $Z=0.007$ the SDU starts at 4$\Msun$, although it is very shallow at this mass.  
The effect of the first and second dredge-up is to lower the C/O ratio from 
its initial value (which we assume is solar, C/O = 0.55) to C/O $\lesssim 0.3$, as 
well as decreasing  the \iso{12}C/\iso{13}C ratio. 

HBB occurs during the TP-AGB when the base of the convective envelope becomes hot enough 
for CNO cycle reactions to occur, and can produce significant increases to the luminosity
and changes to the surface abundances \citep[for more details we refer to reviews by][]{herwig05,karakas14dawes}. 
Of importance for this study is the effect HBB can have on preventing the formation of a carbon-rich
atmosphere, by converting \iso{12}C to \iso{14}N \citep{boothroyd93}. Once HBB ceases, dredge-up
can continue and the star may still become C-rich \citep{frost98a,vanloon99a}. This is observed in
our lowest metallicity models, where all intermediate-mass models become C-rich.
HBB begins to alter the surface composition when the temperature exceeds 
$50 \times 10^{6}$ K (MK), which is reached in models of 4.25$\Msun$ at $Z=0.007$, 
4.5$\Msun$ at $Z=0.014$, and 5$\Msun$ at $Z=0.03$ (Table~\ref{tab:models}).

\citet{karakas14} discussed the effect of helium enrichment on the stellar lifetimes and
core masses in low-mass, low metallicity AGB models. They found that an increase in the 
initial helium abundance leads to a shorter stellar lifetime, as a consequence of less
hydrogen fuel for the main sequence. Stellar lifetimes are reduced in the helium-rich and
metal-rich models, where the total lifetimes decreases by factors of $1.7-2.0$, 
depending on initial mass and helium abundance (see Tables~\ref{tab:models} and~\ref{tab:he}).
The total stellar lifetime of a 3$\Msun$, $Z = 0.03$ model is reduced from 530~Myr when $Y=0.30$
to 283~Myr when $Y=0.40$, a decrease of 47\% and we see similar reductions in the solar metallicity
model of the same mass. AGB lifetimes are reduced by factors of 2.0-3.0, again depending
upon mass and helium composition. The implication here is that AGB stars located in the (helium-rich) 
bulges of galaxies may not be as old as estimated from canonical stellar evolutionary calculations. 

An increase in the core mass on the beginning of the AGB means that the minimum core mass 
for the occurrence of the SDU and HBB  is lowered in helium-rich models \citep{karakas14}. 
The minimum mass for the SDU is reduced from 4$\Msun$ to 3.5$\Msun$ at $Z=0.014$ when 
$Y=0.35$.  Similarly, the minimum mass for HBB is reduced from 4.5$\Msun$ to 4$\Msun$ at 
$Z=0.014$. At $Z=0.03$, the minimum mass for SDU drops from 4.5$\Msun$ to 4$\Msun$ although
the minimum mass for HBB does not change. The effect of HBB on the evolution of the surface 
C/O ratio is weak at the minimum mass; the reduction in third dredge-up efficiency and 
the decrease in the AGB lifetimes are more important, as we discuss next.

\subsection{The third dredge-up}

\begin{figure}
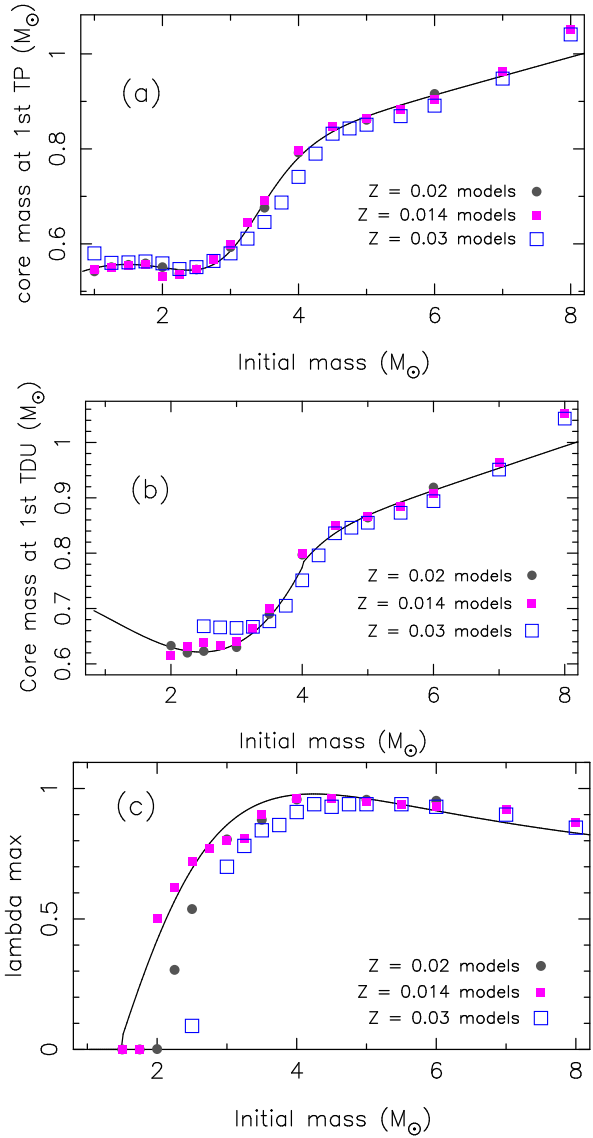

\begin{center}
 \includegraphics[width=5cm,angle=270]{fig1a.ps}
 \includegraphics[width=5cm,angle=270]{fig1b.ps}
 \includegraphics[width=5cm,angle=270]{fig1c.ps}
 \caption{(a) The core mass at the first thermal pulse, $M_{\rm c}(1)$, (b) the core mass
at the  first third dredge-up episode, $M_{\rm c}^{\rm min}$, and (c) the maximum third
dredge-up efficiency parameter, $\lambda_{\rm max}$ for the new $Z=0.014$ (solid
magenta squares) and $Z=0.03$ (large open blue squares) models using data
from Table~\ref{tab:models}. The fits to the $Z=0.02$
models from \citet{karakas02} are shown by the solid lines, as are the model data for 
the $Z=0.02$ models with mass loss (solid dark grey circles), also from \citet{karakas02}.
\label{fig1}}
\end{center}
\end{figure}

\begin{figure}
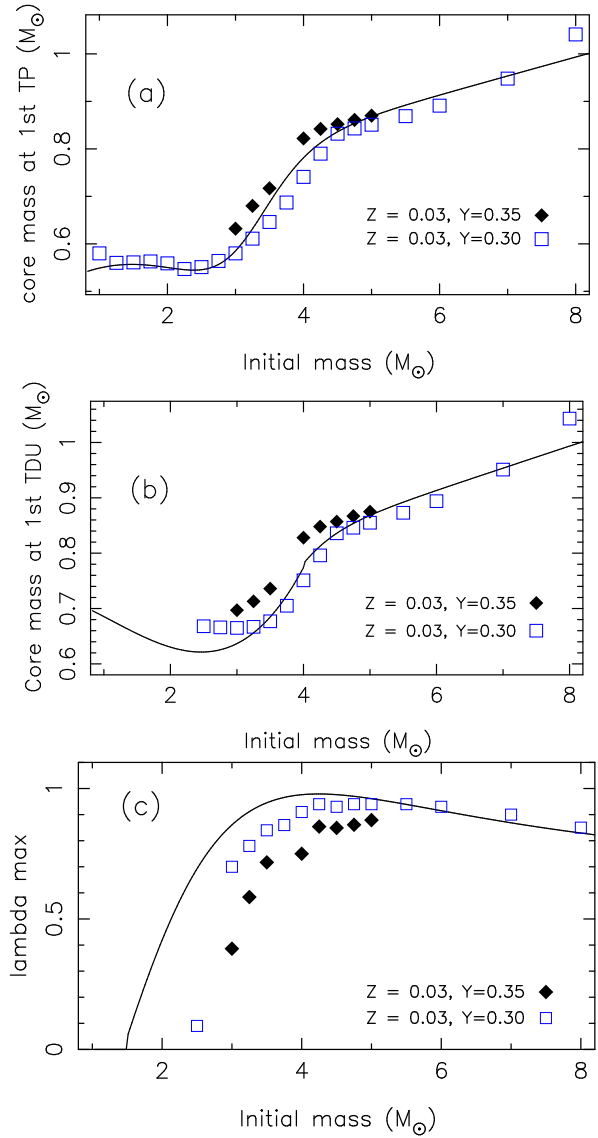

\begin{center}
 \includegraphics[width=5cm,angle=270]{fig2a.ps}
 \includegraphics[width=5cm,angle=270]{fig2b.ps}
 \includegraphics[width=5cm,angle=270]{fig2c.ps}
 \caption{(a) The core mass at the first thermal pulse, $M_{\rm c}(1)$, (b) the core mass
at the  first third dredge-up episode, $M_{\rm c}^{\rm min}$, and (c) the maximum third
dredge-up efficiency parameter, $\lambda_{\rm max}$ for $Z=0.03$ models (large open blue squares) 
with a canonical helium abundance ($Y=0.30$) and for $Z=0.03$ models with $Y=0.35$.
The fits to the $Z=0.02$ models from \citet{karakas02} are shown by the solid lines.
\label{fig2}}
\end{center}
\end{figure}

In Figure~\ref{fig1} we show a comparison between the new $Z=0.014$ and $Z=0.03$ models 
with a canonical helium abundance to the parametrization of the third dredge-up for 
$Z=0.02$ provided by \citet{karakas02}. We also include the model
data for the $Z=0.02$ models with mass loss from that study. The $Z=0.02$ fit is an excellent
match to the core mass at the first thermal pulse for the $Z=0.014$ models and for the $Z=0.03$ 
models for $M \le 3\Msun$. The core mass at the first thermal pulse is smaller in the 
$Z=0.03$ for $3.25 \le M(\Msun) \le 6$ compared to the fit. This mass range experiences 
the deepest third dredge-up, as shown by by Figure~\ref{fig1} (c). 
Similar to the results found by \citet{karakas02}, we find that the core mass at the first thermal
pulse is a good approximation for the core mass at the first TDU episode for $M > 4\Msun$.
The core mass at the first thermal pulse and at the first TDU episode are larger in the
8$\Msun$ models than the parameterization by \citet{karakas02}. This is not entirely surprising
as the fit was made using models with a maximum mass of 6$\Msun$ (although it does do a good
job for the 7$\Msun$ models). 

\citet{karakas02} used solar metallicity models without mass loss to derive the fits
shown in Figure~\ref{fig1}. Our new $Z=0.014$ models are a good match to those fits.
In comparison, the $Z=0.02$ models with mass loss from \citet{karakas02} 
have shallower dredge-up for $M \le 2.5\Msun$.   The $Z=0.03$ models show similar values
to $\lambda_{\rm max}$ as the $Z=0.014$ models for $M \ge 3.5\Msun$; however the lower mass
models at $Z=0.03$ experience considerably shallower dredge-up, with TDU only starting at
2.5$\Msun$ (compared to 2$\Msun$ at $Z=0.014$). While the 2.5$\Msun$, $Z=0.03$ model experiences
some TDU it does not become carbon rich.

The $Z=0.007$ models are similar in metallicity to the $Z=0.008$ models from \citet{karakas02}. 
\citet{doherty14a} present evolution and nucleosynthesis results for a 7$\Msun$, $Z=0.008$ model, 
similar to our most massive case.
The minimum mass for the TDU is 1.5$\Msun$ in the $Z=0.008$ models with mass loss, although
dredge-up is shallow with a maximum $\lambda = 0.084$. This is very similar to the new
$Z=0.007$ models, where dredge-up also starts at 1.5$\Msun$, where $\lambda_{\rm max} = 0.1$.
Dredge-up is deeper in the $Z=0.007$ models for masses between 1.75$\Msun$ and 2.5$\Msun$
but after that the two model sets are similar. The minimum core masses at the first thermal
pulse and first TDU are higher in the $Z=0.007$ models. Besides the difference in metallicity,
the stellar evolutionary code used for the calculations has been updated to use the 
LUNA rate \citep{bemmerer06} for the $^{14}$N(p,$\gamma$)$^{15}$O reaction, which governs the 
main-sequence lifetime and consequently the size of the H-exhausted core at the end. 
Furthermore, the code now uses the NACRE rate \citep{angulo99} for the triple-$\alpha$ 
process and the $^{12}$C($\alpha,\gamma$)$^{16}$O reaction \citep[instead of the rates from][]{cf88}, 
both of which govern energy generation during core He burning and therefore the size of the 
core at the beginning of the AGB \citep[e.g.,][]{castellani92,imbriani01,halabi12}.

In Figure~\ref{fig2} we illustrate the effect of an enhanced helium composition on the
third dredge-up, where we plot the $Z=0.03$ models with $Y=0.35$ against the canonical
helium abundance models. While we only have 8 models for comparison, it is clear that an
enhanced helium abundance increases the core mass at the first thermal pulse, and therefore
at the first TDU episode, and importantly, lowers the maximum third dredge-up
efficiency found at a given mass. From Table~\ref{tab:he} we can see similar behaviour in the 
helium-enhanced $Z=0.014$ models.  The reduction in $\lambda$ is most apparent at 
the lowest masses that experience TDU, which has important consequences for the production 
of carbon stars in metal-rich populations. 

Increasing the helium content changes the mean molecular weight, $\mu$, which results
in hotter H-burning regions and higher luminosities. As a consequence, the whole
star is bigger, brighter and has a lower effective temperature (a consequence of 
a larger radius).  
This means that the mass-loss rate is higher, which in turn reduces the number
of TPs as can be seen most noticeably for the $Y=0.40$ models in Table~\ref{tab:he}.
In the metal-rich models of $Z=0.03$, TDU does not begin until e.g., the 11th TP in the 
3.5$\Msun$ model with $Y=0.30$. The same model of $Y=0.40$ only has 12TP and by the 
11th the total mass has been reduced by $\approx 1\Msun$. This reduction in 
number of TPs may be the main reason for the reduction in $\lambda_{\rm max}$ 
\citep[which has been shown to increase steadily with thermal pulse in lower mass stars, e.g.,][]{karakas02}. 

\subsection{The C/O ratio}

\begin{figure}
\begin{center}
 \includegraphics[width=6cm,angle=270]{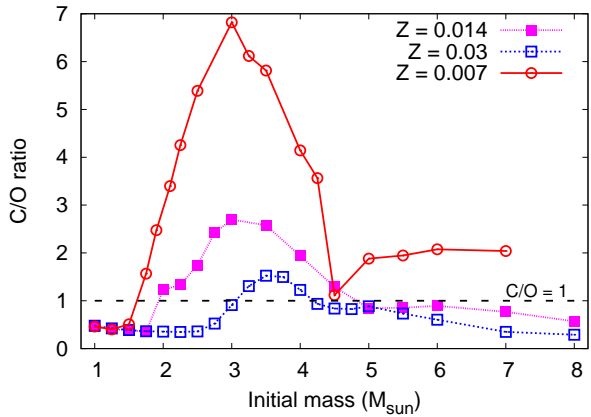}
 \caption{The final C/O ratio at the surface as a function of initial stellar mass for
the stellar models in Table~\ref{tab:models}.
\label{fig3}}
\end{center}
\end{figure}

In Figure~\ref{fig3} we show the range of final C/O ratios predicted from the canonical
stellar evolutionary sequences. All $Z=0.007$ models more massive than 1.5$\Msun$ become
C-rich by the tip of the AGB, including the intermediate-mass AGB stars that suffer HBB.
The final C/O ratio does not reflect the evolution of the C/O ratio for HBB models,
which spend the majority of the TP-AGB with C/O $< 1$ as a consequence of efficient envelope
burning. As an example, the 6$\Msun$, $Z=0.007$ model experiences 64 TPs and only becomes C-rich
after the 62nd, from Table~\ref{tab:models} we see that the C-rich phase lasts for about
2\% of the entire AGB phase.   At higher metallicities dredge-up also occurs
but there is less oxygen depletion by HBB (and the initial O abundance is much higher) which
means that the models do not become C-rich.

Lower mass stars that do not experience HBB can have C-rich lifetimes that are 
$\gtrsim 10$\% of the total AGB lifetime in the lowest metallicity models (Table~\ref{tab:models}).
The ratio of the C-rich lifetime to the total AGB lifetime can be used as a proxy for 
the ratio of the number of C stars to O-rich M-type AGB stars, C/M, in a given population.
In the metal-rich models of $Z=0.03$ this ratio has a maximum of $\approx 0.03$ at 3.75$\Msun$,
whereas it reaches much higher values of $\approx 0.07$ in solar metallicity models.
Increasing the helium content of the model greatly reduces the number of carbon stars,
where Table~\ref{tab:he} shows that an increase of $\Delta Y = 0.07$ at solar metallicity
reduces the ratio to 0.02. In the metal-rich models an increase of helium almost entirely
wipes out the C-star population. 

Figure~\ref{fig3} illustrates that the mass range of carbon-stars is predicted to
shrink with an increase in the metallicity. By the time we get to $Z=0.03$ only a narrow
mass range between $3.25\Msun$ to 4$\Msun$ become carbon rich, compared to 
2$\Msun$ to 4.5$\Msun$ at $Z = 0.014$. Furthermore, the maximum C/O ratio also decreases
with increasing metallicity as a consequence of a higher initial oxygen abundance.
Models with masses near 5$\Msun$ at $Z=0.014$ and $Z=0.03$ experience mild
HBB. It is enough to keep the C/O ratio less than unity but Figure~\ref{fig3} shows
that the final C/O ratio is $\approx 0.9$. One or more TDU episode (at the same
efficiency) would be enough for the model to become C-rich.

All intermediate-mass stars experienced convergence problems 
of the type discussed by \citet{lau12} and the models terminated with reasonably large
envelope masses ($\approx 1\Msun$). This means that the lifetimes given in Tables~\ref{tab:models}
and~\ref{tab:he} are lower limits although in \citet{karakas14} we estimated that one or two
missed TPs is not going to affect the total stellar lifetimes or AGB lifetimes of 
models with $M \lesssim 2.4\Msun$.  For intermediate-mass AGB stars, we may be missing
up to 3 or more TPs. As an example, we estimate that the 5$\Msun$, $Z = 0.03$ model
may experience another 3 TPs, which would increase the AGB lifetime by $\approx 33,000$
years (taking a maximum interpulse period of 11,000 years). This is a negligible
increase to the total and AGB lifetimes but increases the TP-AGB phase by about 16\%.

\begin{figure}
\begin{center}
 \includegraphics[width=6cm,angle=270]{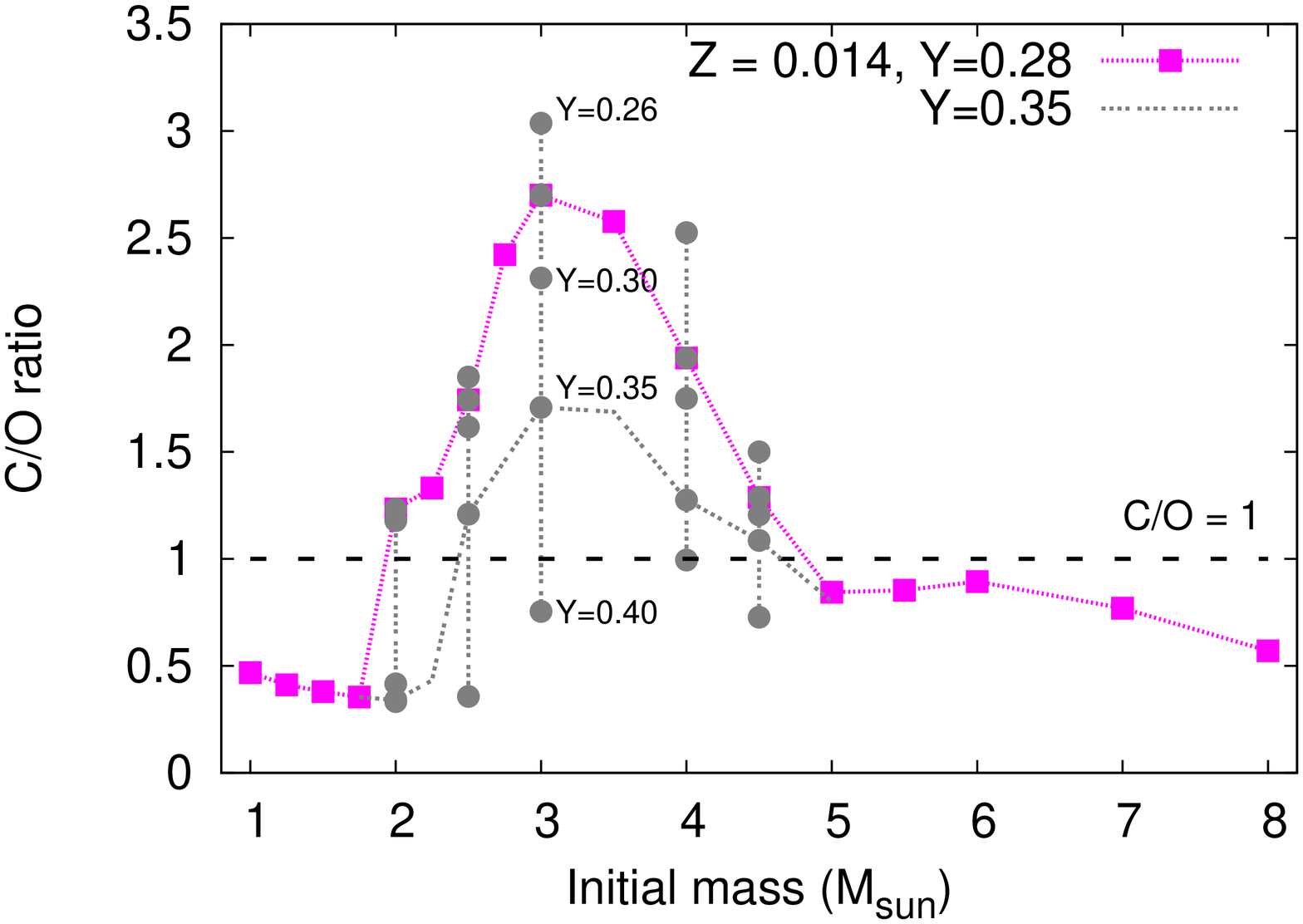}
 \includegraphics[width=6cm,angle=270]{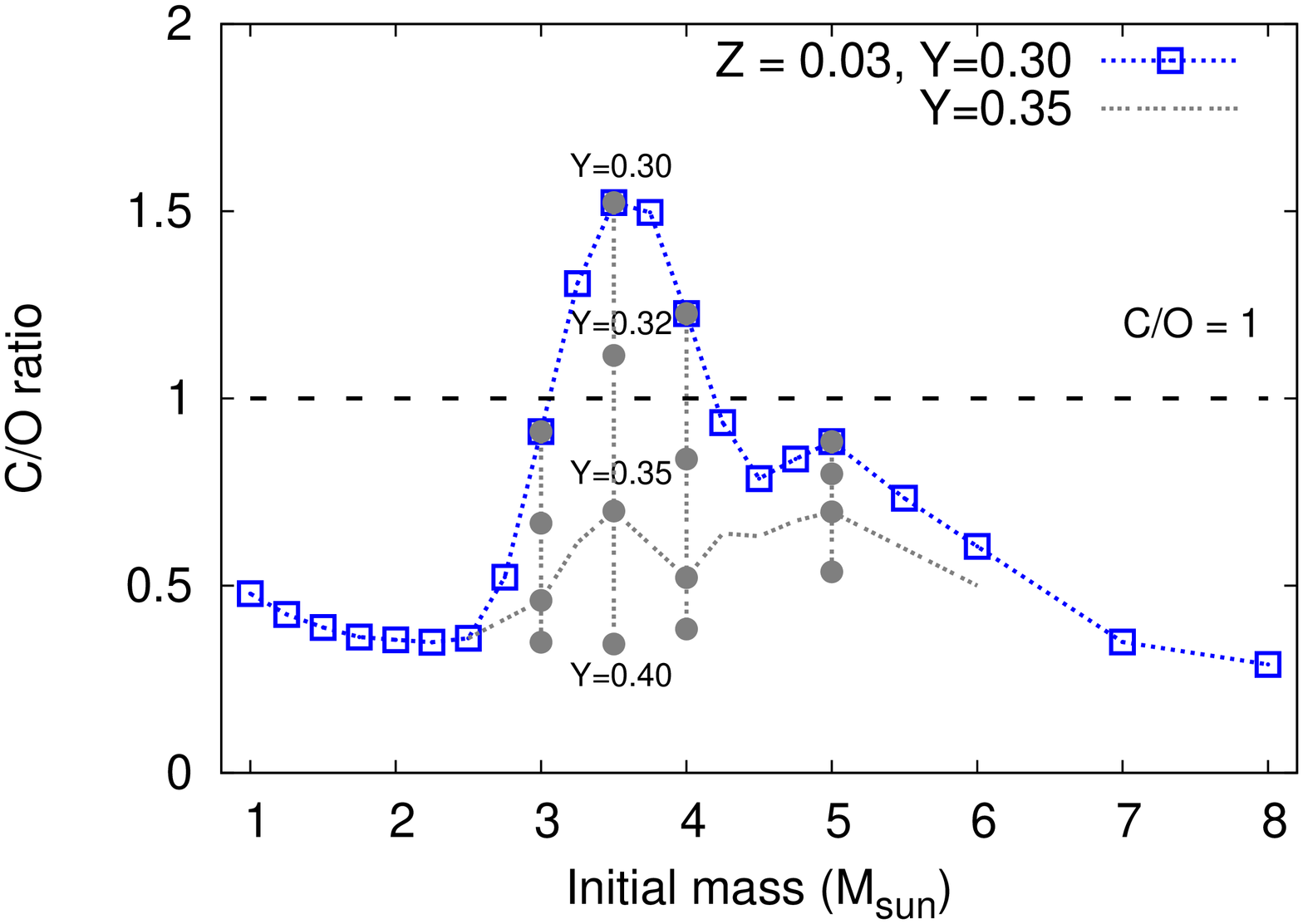}
 \caption{The final C/O ratio for the (a) $Z = 0.014$ models, including the models with
a helium compositions between $Y=0.26$ to $Y=0.40$, and (b) the $Z = 0.03$ models, including models
with helium compositions between $Y=0.32$ to $Y=0.40$.
\label{fig4}}
\end{center}
\end{figure}

In Figure~\ref{fig4} we show the final C/O ratios for the stellar models calculated with different
helium compositions. We draw a line through the points with $Y=0.35$ to highlight how the 
C/O ratio decreases in helium-rich AGB models.  The main point to take away from this diagram is that
an increase in helium by only $\Delta Y = 0.05$ is enough to inhibit carbon star production
altogether at $Z=0.03$. Models that experience the largest shift toward lower C/O ratios
are those at the minimum mass for the onset of the TDU, that is, stars with $M\approx 2-3\Msun$
according to our models. 

\subsection{Comparison to other studies} \label{sec:compare}

While there are no other $Z=0.03$ AGB models for comparison, we can compare the solar and $Z=0.007$
models to other studies.   Here we compare to the 2$\Msun$, $Z = 0.0138$ model from
\citet{cristallo09} and to the 3$\Msun$, $Z = 0.02$ model from \citet[][noting that 
evolution model data is not available in the paper for the 3$\Msun$, $Z=0.0138$ model]{cristallo11}.
We also provide a comparison between our most massive AGB model of 7$\Msun$, $Z=0.007$
to the calculations by \citet{ventura13}, which use the Full Spectrum of Turbulence convective
prescription. While there are other AGB models for comparison, including the intermediate-mass
and super-AGB AGB models by \citet{doherty14a} and the MESA models by \citet{pignatari13}, we 
limit our comparison for the sake of brevity.

Our 2$\Msun$, $Z = 0.014$ model becomes carbon rich, with a final C/O = 1.23 and dredges
up a total of 0.0236$\Msun$ over 25~TPs during the TP-AGB. In comparison, the 2$\Msun$, $Z=0.0138$ 
model from \citet{cristallo09} experiences 12~TPs during the AGB and dredges up a total of 
0.0362$\Msun$, with a final C/O=1.88.
The model by Cristallo et al. also becomes C-rich at a lower core mass of 0.603$\Msun$ compared
to our 0.65$\Msun$ (and presumably at a lower luminosity, although this is not included 
in their tables). 
 
The model by \citet{cristallo09} has convective boundary mixing included, through the use
of an exponentially-decaying diffusive overshoot scheme dependent on the parameter $\beta$,
which is similar to the scheme adopted by \citet{herwig00}. While our model has no
formal mixing beyond the Schwarzschild border, we adopt the search for a neutral border to the 
convective-radiative boundary described by \citet{lattanzio86} which has been found to 
increase the amount of TDU relative to schemes that strictly adopt the Schwarzschild criterion 
\citep{frost96}. Adopting the mixing scheme used by \citet{cristallo09} with their choice of 
$\beta$, leads to deeper TDU at a much lower core mass than we find in our calculation 
\citep[see also discussion in][]{herwig00}.  We would need to include such a mixing scheme if
we were to try and reproduce the Galactic C-star luminosity function, which peaks at a
bolometric luminosity of about  $-4.9$, where there are very few Galactic C-stars with luminosities higher 
than this \citep{whitelock06,guandalini06,guandalini13}. 

We can include convective overshoot in a simple manner by extending the base of the convective
envelope by $N$ pressure scale heights, as done by \citet*{karakas10b}. If we set $N=2$, we can
calculate a 2$\Msun$, $Z=0.014$ AGB model with similar characteristics to the model by \citet{cristallo09}.
The minimum core mass for TDU is reduced from 0.616$\Msun$ with no overshoot
to 0.577$\Msun$, which is similar although still slightly higher than the minimum core
mass for TDU of 0.568$\Msun$ found by \citet{cristallo09}. 

We now compare the 3$\Msun$, $Z=0.014$ model to the 3$\Msun$, $Z=0.02$ model by \citet{cristallo11}.
The 3$\Msun$ model by Cristallo et al. enters the AGB at a much higher core mass of 0.653$\Msun$
and becomes C-rich at a core mass of 0.677$\Msun$. The final core mass is 0.700$\Msun$. In comparison,
the slightly lower metallicity Stromlo model enters the AGB at a core mass of 0.598$\Msun$, becomes
C-rich at 0.669$\Msun$ and has a final core mass of 0.691$\Msun$. The Stromlo model experiences 
28~TP, and has a final C/O=2.7, and dredges up 0.1$\Msun$ from the He-shell. In comparison, the
Cristallo et al. model has 14 TPs, a final C/O=1.59, and dredges up 0.0754$\Msun$. 
The C-rich lifetime of the  \citet{cristallo11} model is shorter, at 0.462~Myr relative to our
0.604~Myr. Increasing the mass-loss rate on the AGB in our calculation would reduce the 
difference in final C/O and C-rich lifetimes.  In summary, the Cristallo et al. model has a 
higher core mass and presumably a higher luminosity, although within the range of uncertainties in 
stellar models the results are reasonably consistent. 

While both the current set of models and models by \citet{cristallo11} predict the occurrence
of carbon stars with masses above 2$\Msun$, we emphasise that C-stars evolving from such progenitor 
stars will be very rare in stellar populations for the following reasons: 1) the initial mass function 
favours stars of lower mass, and 2) because they have short AGB and C-rich lifetimes.

We now compare the 7$\Msun$, $Z=0.007$ model to the 7$\Msun$, $Z=0.008$ model by
\citet{ventura13}. This model was calculated with the Full Spectrum of Turbulence convective
prescription which results in a stronger HBB during the AGB \citep{ventura05a}. This means
that the peak HBB temperature is higher, at 105~MK, than the Stromlo model which peaks
at 103~MK (noting that the Stromlo model has a lower metallicity, which also results in 
higher temperatures). The Stromlo model becomes C-rich at the very tip of the AGB, has a final
helium mass fraction at the surface of $Y=0.352$, and experiences 61 TPs. The final core
mass is 1.04$\Msun$. In comparison, the 7$\Msun$ model by Ventura et al. only has 24~TP,
destroys considerable carbon such that the final C/O is well below unity, has a final
surface $Y=0.36$, and a final core mass of 1.14$\Msun$. 

The 7$\Msun$ Stromlo model becomes carbon rich because TDU continues after
the cessation of HBB, which allows the C abundance to increase \citep{frost98a}.
\citet{vanloon99a} presented observational evidence that supports this scenario, 
finding a sample of very luminous, dust-obscured C-rich AGB stars in the Magellanic Clouds. 
The existence of very bright, C-rich AGB stars is also evidence that stars in this mass 
range experience TDU at the metallicities of the Magellanic Clouds.

\section{Modelling uncertainties} \label{sec:uncert}

The evolution and nucleosynthesis of low-mass and intermediate-mass stars is significantly
affected by numerical modelling uncertainties as well as uncertainties in the input physics
\citep[see][for a detailed discussion]{karakas14dawes}. The main uncertainties affecting the current
study are the treatment of convection and in particular, the numerical treatment of convective
borders and the mass-loss rate used on the AGB 
\citep[e.g.,][]{ventura05a,ventura05b,stancliffe07a,cristallo09}. The method for determining
convective borders in particular determines the occurrence of the TDU \citep{frost96,mowlavi99a}, 
and the minimum initial stellar mass for carbon-star production, while the mass-loss rate 
determines the number of thermal pulses and the AGB lifetime \citep{bloecker95}.

For models near the minimum mass for the onset of the TDU, dredge-up becomes 
less efficient as the metallicity increases. The treatment of convection and of 
convective borders is of paramount importance here. Convective boundary mixing as applied 
by e.g, \citet{herwig00}, will decrease the minimum mass for carbon star production to whatever 
mass one desires. However, observations of C stars in the Galaxy are hindered by uncertain 
distances which means that the minimum mass from observations is not well constrained 
\citep{wallerstein98}, although can be inferred from carbon-star luminosity functions to
be $\approx 1.5\Msun$ \citep[e.g.,][]{whitelock06,guandalini06,guandalini13}. 

\citet{guandalini13} re-derived the C-star luminosity function for Galactic C stars and find
good agreement with the theoretical C-star luminosity function from \citet{cristallo11}, which
has a minimum C-star mass of 1.5$\Msun$ at solar metallicity. As described earlier, the models 
by \citet{cristallo11} employ a convective boundary mixing scheme, which requires a free parameter 
in order to obtain dredge-up at the lowest masses.
We require a considerable amount of convective overshoot (3 pressure scale heights) for the 
1.5$\Msun$, $Z=0.014$ model to become C-rich. While large, this is similar to the amount of 
overshoot \citet{kamath12} required in order to  get the correct O-rich to C-rich 
transition luminosity in lower metallicity Magellanic Cloud clusters. 
Note that a similar amount of overshoot lowers the minimum mass for 
C-star production from 3.25$\Msun$ at $Z=0.03$ to 2.5$\Msun$.

The models by \citet{cristallo11} and \citet{ventura13} experience many fewer TPs than the calculations
presented here. While the convective prescription in the envelope plays a more dominant role in 
intermediate-mass models \citep{ventura05a}, in lower-mass models the mass-loss rate
on the AGB is crucial. We use the \citet{vw93} semi-empirical prescription, which was derived
from a sample of Galactic and Magellanic Cloud O and C-rich AGB stars. The mass-loss rate
used by \citet{cristallo09} is based on a similar semi-empirical calibration to \citet{vw93} 
of the period-mass loss relations of long-period variables \citep{straniero06}.

Star clusters in the Galaxy can be used to probe uncertain physics in stellar evolutionary
calculations such as mass loss and convection 
\citep{weidemann00,marigo01,ferrario05,kalirai08,kalirai09,cristallo11}. The core mass
at the first thermal pulse is a good estimate of the final mass for stars more massive than about
4$\Msun$, which only experience shallow core growth owing to efficient TDU and short interpulse
periods during which the H-burning shell is dominant. For stars less massive than about 4$\Msun$,
the core growth is more significant and is determined by the depth of TDU 
and the mass-loss rate, which determines the AGB lifetime and hence the amount
of core growth \citep[e.g.,][]{kalirai14}. 

\citet{kalirai07} determined the initial masses of the white dwarfs ($0.61 \pm 0.02\Msun$) 
in the solar metallicity cluster NGC~7789 to be $2.02\pm 0.07\Msun$. Our 2$\Msun$, $Z=0.014$
model has a final mass of 0.659$\Msun$. In order to reproduce a final mass of 0.61$\Msun$,
we require convective overshoot on the order of 2 pressure scale heights from the formal
border. Again, this is consistent with previous studies and suggests that convective
boundary mixing occurs in such stars.  

The cluster NGC 6791 has a metallicity of [Fe/H] = $+0.4$,
similar to our $Z=0.03$ models. However, our 1$\Msun$, $Z =0.03$ produces a final core 
mass of 0.57$\Msun$, larger than the mass of the white dwarfs in this cluster, at 0.53$\Msun$
\citep{kalirai08}. At such low initial masses, the TDU is not predicted to occur so the uncertain details of 
He-burning determine the size of the core on the AGB \citep[see discussion in][]{karakas14dawes}.
Mass loss on the RGB is also important, as discussed by \citet{kalirai07}. 
Helium enrichment increases the core masses of the models at the start of the AGB, which
implies that it is unlikely that any of the cluster member stars are strongly helium rich.

\citet{kalirai14} compared the core mass growth from theoretical calculations using different
mass-loss prescriptions to the white dwarf masses in Galactic open clusters of solar and 
super-solar metallicity.  These authors found that the \citet{vw93} prescription agrees very 
well with the observational data, at least for AGB stars up to about 2$\Msun$.
In comparison, comparisons between the observational data and results using the
\citet{bloecker95} rate with $\eta = 0.2$ and the \citet{vanloon05} rates were not so favourable.

For masses above 2$\Msun$, it becomes harder to constrain the mass-loss rates of AGB stars because
of the paucity of observational data. A comparison between our 3$\Msun$, $Z=0.014$ model to 
a similar metallicity calculation by \citet{cristallo11} suggests that the \citet{vw93} mass-loss
formula is too low at these masses.  \citet{ventura00} calibrated the \citet{bloecker95}
rate against the lithium-rich, O-rich AGB stars in the Magellanic Clouds and determined a value
of $\eta = 0.01$. 

We test the \citet{bloecker95} rate with $\eta = 0.01$ in a 3$\Msun$, $Z=0.03$,
$Y=0.30$ model and obtain a much lower mass loss relative to the \citet{vw93} prescription.
The 3$\Msun$, $Z=0.03$ model now has 47 TPs relative to 33, and becomes C-rich, with a final
C/O = 1.7 whereas with \citet{vw93} the model does not become C-rich. We find that the 
\citet{bloecker95} mass-loss rate with $\eta = 0.03$ results in a similar number of TPs
to the \citet{vw93} mass-loss rate at 3$\Msun$, $Z=0.03$. Setting $\eta = 0.1$ results
in a mass-loss rate that is so fast that all the envelope is lost before TDU begins.
Setting $\eta = 0.01$ in a 3$\Msun$, $Z=0.03$ model with $Y=0.35$ results in the formation
of C star, where the final C/O=1.35. This indicates that the amount of helium needed
to remove carbon stars from a population is dependent on the mass-loss rate.

The initial solar abundances adopted are not a significant uncertainty on the stellar
evolutionary calculations of solar metallicity. This is demonstrated by the reasonably good
agreement between the $Z=0.014$ models presented here and the previous $Z=0.02$ models.
To test this assumption further, we adopt the solar abundances by \citet{lodders09}, 
which have a proto-solar metallicity of $Z=0.0153$, in a model of 3$\Msun$. The 3$\Msun$, $Z=0.0153$
model has very similar characteristics to the model of $Z=0.014$:
29~TPs instead of 28, 0.091$\Msun$ of material dredged up relative to 0.1$\Msun$, and the
same tip AGB luminosity of $13,000\Lsun$.

Finally, we comment on other uncertainties that affect models of low and intermediate-mass
stars including non-convective extra mixing and stellar rotation. While non-convective extra mixing
on the RGB seems ubiquitous in all low-mass stars below 2$\Msun$
\citep[e.g.,][]{gilroy89,eggleton08,charbonnel10}, extra mixing on the AGB is much less 
certain \citep{busso10,karakas10b}. By extra mixing in this context, we are referring
to the mixing between the base of the convective envelope and hydrogen shell, such that
the products of H-burning are observed at the surface. 

The mechanism(s) responsible for mixing material from the base of the convective envelope
through a hot region near the H-shell is unknown. Various mechanisms have been proposed
including including thermohaline mixing (described below), magnetic buoyancy \citep{nordhaus08,nucci14},
and stellar rotation \citep{herwig03,piersanti13}. Rotation can generally be ruled out for 
RGB stars \citep[e.g.,][]{palacios06,charbonnel10}, although it can produce changes to 
AGB nucleosynthesis \citep{herwig03,piersanti13}. From the models of \citet{piersanti13}, rotation
does not appear to limit the production of C-stars at a given mass at solar metallicity
although it does reduce the final [C/Fe] at the surface.

Thermohaline mixing or ``double-diffusive mixing'' has been shown to be effective on the RGB 
\citep{charbonnel07a,eggleton08,denissenkov10,angelou12} but less so on the AGB 
\citep{stancliffe09,stancliffe10}. Thermohaline mixing has been coupled with magnetic 
fields \citep{denissenkov09} and with rotation \citep{lagarde11}. 

The main effect of extra mixing is to lower the \iso{12}C/\iso{13}C ratio, with only 
small changes to the C/O ratio. For example, \citet{lederer09b} observed C/O=0.2 in 
unevolved AGB stars in the LMC cluster NGC~1978, about 0.1 below the predicted C/O=0.30. 
Extra mixing does not appear to be operating in the C-rich AGB stars in NGC~1978
but it does appear to be moderately efficient in the AGB envelope of the LMC cluster NGC 1846, which
has a similar metallicity to NGC 1978 \citep{lebzelter08}. The operation of extra 
mixing in low-mass AGB stars does not change their C-rich status, but instead acts to
lower their \iso{12}C/\iso{13}C ratios relative to models without extra mixing. 

\section{Conclusions}  \label{sec:conclude}

We present grids of new stellar evolutionary models with masses between 1$\Msun$ to 7-8$\Msun$
at metallicities of $Z=0.007$, 0.014, and 0.03. These metallicities
are appropriate for comparison to AGB and PNe populations in the disc and bulge of the 
Milky Way Galaxy and external galaxies such as M31 and the LMC.
In a future study, we will calculate detailed nucleosynthesis predictions from these models
in order to produce stellar yields and for comparison to the abundances of observed objects.

The $Z=0.03$ models are the first detailed AGB models of this metallicity in the 
literature. We compare our results to the parameterization of the TDU
by \citet{karakas02} for AGB models of $Z=0.02$. These fits were made to models
without mass loss and were calculated with a higher initial helium composition of
$Y =0.2928$. The $Z=0.02$ models with mass loss experience shallower dredge-up than
we find in the $Z=0.014$ models with $Y=0.28$ but are much closer to what we find
for the $Z=0.014$ models with $Y=0.30$. 

The behaviour of the TDU in the canonical $Z= 0.014$ models is well approximated by 
the parameterization for $Z=0.02$ by \citet{karakas02}. In contrast, the $Z=0.03$ models 
experience considerably shallower dredge-up for $M < 4\Msun$ compared to the solar
metallicity models. The mass range that produces carbon stars is 1.75--7$\Msun$ at $Z=0.007$,
 2--4.5$\Msun$ at $Z=0.014$,   which is reduced to 3.25--4$\Msun$ at $Z=0.03$. 
The 3$\Msun$, $Z=0.03$ model almost becomes C-rich, where the final C/O $=0.91$.
We have discussed how uncertain input physics such as the AGB mass-loss rate and
the treatment of the border between convective and radiative regions will impact
the calculations. Other uncertainties such as the molecular opacities \citep[e.g.,][]{marigo02} 
can also have a considerable impact on AGB lifetimes and will shift the minimum mass
for C-star production.

In this study we also present the first helium-rich stellar evolutionary models at a solar
and super-solar metallicity. We find that the third dredge-up is either reduced or inhibited
when the initial helium content of the model is increased. This is caused by a reduced number
of thermal pulses on the AGB relative to the canonical models. A small increase of
$\Delta Y = 0.05$ is enough to inhibit carbon star production altogether at $Z=0.03$, depending
on the choice of mass-loss rate, whereas at solar metallicity we require a much larger 
helium enrichment of $\Delta Y\approx 0.1$ to prevent the formation of carbon stars. 
The consequences of removing  carbon stars from super-solar metallicities on the 
chemical evolution of galaxies still needs to be explored, once stellar yields 
from such models become available.

In the inner region of M31, \citet{boyer13} found a very low number of carbon stars  
compared to the fit through observations from nearby galaxies covering a range
of metallicities. The fit suggest there  should be a much higher C/M star fraction at 
[M/H] $\approx 0.1 \pm 0.1$ (their Figure 1). The fit is also consistent with
model predictions for close to solar metallicity such as those presented here,
which suggests that there should be a reasonable number of carbon stars in the
inner region of M31.   Indeed, if anything we are underestimating the number of 
carbon stars because we only find substantial dredge-up for $M \ge 2\Msun$.
The lack of C-stars is not simply because \citet{boyer13} are sampling an older 
population, where optical colour-magnitude diagrams for their region reveal the
presence of stellar populations with a turn-off age younger than $\approx 1$~Gyr
\citep{boyer13}.  We conclude instead that the lack of carbon stars in the inner region of M31 is
either the result of the metallicity being higher than estimated, at [Fe/H] 
$\approx 0.3$ instead of 0.1, and/or there is a substantial enrichment in helium
in the observed stellar population. We show in Table~\ref{tab:he} that an increase of 
$\Delta Y \approx 0.1$ at solar metallicity removes all of the carbon stars from the
population.

Finally, helium appears to be an important third parameter governing the evolution and
nucleosynthesis of low and intermediate-mass AGB stars. We have shown that helium is as important
for the evolution of AGB stars at solar and super-solar metallicities as it is in low
metallicity AGB stars \citep{karakas14}.   We speculate that
it may be possible to take the ratio of C/M stars observed in e.g., M31 by \citet{boyer13} 
and infer the level of helium enrichment in the inner regions of galaxies if the metallicity
is well determined.  Strong levels of
helium enrichment reduce or completely remove carbon stars from populations. 
Estimating helium abundances from C/M star ratios would certainly  be a novel way to 
infer the chemical enrichment of galaxies.

\section*{Acknowledgments}

The author thanks the anonymous referee for comments that have improved the manuscript.
The author also thanks Melissa Ness and David Nataf for comments on the manuscript, and 
Henry Poetrodjojo for calculating a two of the helium-rich stellar evolutionary models used 
for this study.
AIK was supported through an Australian Research Council Future Fellowship (FT110100475).

\bibliographystyle{apj}
\bibliography{mnemonic,library}

\bsp

\label{lastpage}

\end{document}